\documentclass[trackchanges,twocolumn]{aastex62}
\usepackage{amsmath,amstext}
\usepackage{units}
\usepackage{url}
\usepackage{xspace}

\newcommand{\bi}{\begin{itemize}}
\newcommand{\ei}{\end{itemize}}
\newcommand{\commentOut}[1]{}

\shortauthors{Lee \& Shen}

\begin{document}

\title{\bf \Large{A Single Binary May Host Recurrent Thermonuclear Supernovae}}

\author{Kaela J.\ Lee}
\affiliation{Department of Astronomy and Theoretical Astrophysics Center, University of California, Berkeley, CA 94720, USA}

\author[0000-0002-9632-6106]{Ken J.\ Shen}
\affiliation{Department of Astronomy and Theoretical Astrophysics Center, University of California, Berkeley, CA 94720, USA}

\correspondingauthor{Kaela J.\ Lee}
\email{kaelajlee@berkeley.edu}

\begin{abstract}

The most commonly accepted progenitor system for Type Iax supernovae (SNe~Iax) is the partial deflagration of a near-Chandrasekhar-mass white dwarf (WD) accreting from a non-degenerate helium donor star, leaving a bound remnant following the explosion. In this paper, we investigate whether the WD remant can  undergo multiple SNe during the system's lifetime. We use Modules for Experiments in Astrophysics (MESA) to evolve various single-degenerate binaries to determine which could plausibly undergo multiple SNe~Iax due to multiple helium accretion phases. We also investigate the possibility for a subsequent Type Ia SN after the formation of a double WD system. Our work concludes that close binaries with relatively high-mass donors produce the highest probability for several thermonuclear SNe. 
\\

\end{abstract}


\section{Introduction}

Type Iax supernovae (SNe~Iax; also referred to as 2002cx-like based on their prototypical member) are explosive transients similar in many ways to Type Ia supernovae (SNe~Ia). Both are believed to be thermonuclear explosions of white dwarfs (WDs) and share spectral similarities such as a lack of hydrogen lines and the presence of intermediate-mass and iron-group elements \citep{Foley2013, Jha2017}. It is estimated that SNe~Iax occur at approximately 5-30\% the rate of SNe~Ia \citep{Foley2013, Liu2015}. Like SNe~Ia, there exists a correlation between luminosity and light curve shape for SNe Iax; however, the correlation for SNe~Iax is offset with increased scatter \citep{Foley2013}. In comparison to SNe~Ia, characteristics of SNe~Iax include lower peak luminosities, lower maximum velocities, lower ejecta masses \citep{Foley2013, Takaro2020}, and unique late-time spectra \citep{Li03,Foley2016}. While SNe~Ia are found in both early- and late-type galaxies, all SNe~Iax have been found in late-type galaxies with the exception of SN 2008ge \citep{Foley10,Takaro2020}. The presence of SNe~Iax in young host galaxies suggests that progenitor systems must produce and explode WDs with short delay times between star formation and supernova. The only thermonuclear explosion for which a potential companion star was observed before the explosion is SN~2012Z, whose companion was a luminous blue star with a mass of $\sim 2 \, M_\odot$ \citep{McCully2014}.

Although the SNe~Iax class is relatively new, there has already been significant progress in narrowing down potential progenitor scenarios. The current leading contender involves the partial deflagration of a WD with a mass near the Chandrasekhar limit ($M_\textrm{Ch}$) accreting from a helium donor star through Roche lobe overflow (but see, e.g., \citealt{kash18a} and \citealt{bobr22a} for alternative scenarios). \cite{IbenTutukov1994} first suggested that a degenerate star accreting from a helium donor could potentially reach $M_\textrm{Ch}$ and produce a supernova; subsequent calculations by \cite{Yoon03} showed that this evolution was indeed plausible. The helium star donor is necessary to explain the short delay times in a Chandrasekhar-mass scenario since the stable mass transfer rate of helium is higher than for hydrogen \citep{Ruiter2009,Jha2017}. Deflagration rather than detonation is thought to explain the lower kinetic energies and greater chemical mixing of the ejecta \citep{Foley2013}. We also note that thermonuclear SNe require the ignition of carbon in the core;  by contrast, off-center ignition results in stable burning and the formation of a massive O/Ne WD or a neutron star \citep{Shen12,Schwab2016}. The deflagration is considered ``partial'' if the destruction of the WD is incomplete. In fact, a study by \cite{Jha2017} concluded that since the final velocities of the ejecta are typically lower than the escape velocity from the surface of the WD, some WD material likely remains bound. Additionally, studies of three-dimensional hydrodynamic simulations of the deflagration model also suggest a bound remnant is left behind \citep{jord12a,Michael2014, Min2014,Lach22}. This leads to the natural question of whether the surviving remnant could begin a subsequent accretion phase through Roche lobe overflow and undergo multiple SNe~Iax. 

In section \ref{sec:mod}, we discuss the methods used in this investigation to determine the viability of multiple SNe Iax. In section \ref{sec:results}, we discuss the results of our simulations and describe the evolution of several systems in detail. We conclude in section \ref{sec:conc}.


\section{Description of models and calculations}
\label{sec:mod}

We model the evolution of a variety of single-degenerate binary systems using Modules for Experiments in Stellar Astrophysics (MESA) \citep{paxt11,paxt13,paxt15a,paxt18a,paxt19a}. The donor star for each of our systems starts as a helium main sequence star. The white dwarf (WD) accretor is modeled as a point source. When the WD accretes up to the Chandrasekhar limit ($M_\textrm{Ch}$), we simulate a Type Iax event by decreasing the WD mass and adjusting the binary separation based on the angular momentum loss due to the ejected mass. We do not alter the mass or structure of the helium star, which is a reasonable assumption for compact donors, but is less so for helium giants due to the impact of the SN ejecta \citep{Pan12,Pan13,Liu13,Bauer19}.  We examine which systems begin mass accretion again and bring the remnant back up to $M_\textrm{Ch}$. Additionally, each binary eventually becomes a double WD system with the potential to undergo a SN~Ia by in-spiralling due to gravitational wave emission \citep{Guillochon2010,Dan2012}. We vary two main initial parameters: the helium star's initial mass and the initial period of the system. We keep the initial mass of the white dwarf fixed at $1.0 \, M_\odot$ for this investigation. Notably, studies by \cite{Wong2019} and \cite{Wong2021}  performed similar simulations using MESA to investigate thermonuclear explosions of helium star and WD binaries. However, our investigation differs by focusing on the post-explosion evolution of these systems.

We begin by using MESA to create helium main sequence star models with masses of $1.0$, $1.5$, and $2.0 \, M_\odot$. Details for using MESA to create helium stars are located in the Appendix. We set the chemical composition to be 98\% $^4$He and 2\% $^{14}$N, by mass; we assume nitrogen is the endpoint of the CNO isotopes after CNO burning. For the initial masses we consider, the helium star expands  after it exhausts helium in the core, potentially leading to mass transfer due to Roche lobe overflow.

After performing the same stellar evolution with four nuclear reaction networks within MESA (\texttt{basic}, \texttt{approx21}, \texttt{co\_burn}, and \texttt{mesa\_201}), we determine that the mass fractions for the most abundant isotopes ($^4$He, $^{12}$C, $^{14}$N, and $^{16}$O) are consistent across all networks within a few percent. Additionally, these various networks do not noticeably alter the stellar evolution of the models. We thus choose the \texttt{basic} network for simplicity.

We then evolve the previously created helium star models in a binary system with a $1.0 \, M_\odot$ point mass that is implied to be a C/O WD.  It has been suggested that more massive WDs that have undergone previous carbon-burning may have a hybrid structure with an O/Ne mantle and a C/O core \citep{deni13b}, which could still allow for central carbon ignition and an eventual SN~Iax (e.g., \citealt{meng14a,krom15a,wilc16a}); however, for simplicity and because WDs with such structures may not exist \citep{leco16a,broo17a,schw19a}, we implicitly assume an initially lower-mass C/O WD accretor.  For each helium star mass, we vary the initial period of the system. During our process, some MESA settings differed between systems in order to allow them to run more smoothly. Details are located in the Appendix.

In general, Roche lobe overflow can occur due to stellar expansion or shrinking binary separation due to loss of angular momentum. In our investigation, we turn off angular momentum loss from magnetic braking since our helium main sequence stars do not have large convective zones at their surface. We do allow angular momentum loss due to gravitational wave radiation. However, since the main sequence lifetime for helium stars of this mass is short (on the order $10^7$ years), the probability of encountering systems with small initial binary separations such that mass transfer begins during the main sequence phase is low.  

The smallest initial period for each donor star mass is set such that the Roche lobe radius is slightly larger than the helium star's radius. Additional systems were constructed with periods increasing by factors of approximately 2. The WD is approximated as a point mass with perfectly conservative mass transfer, which allows transfer to occur both below and above the thermally stable WD accretion rates \citep{Brooks2016}. The limitation of simulating the WD as a point source is the inability to determine whether carbon ignition begins centrally or off-center, which determines whether a supernova occurs. By assuming conservative mass transfer, we maximize the total mass and orbital angular momentum of the system. The results in this work can thus be regarded as maximizing the number of SNe~Iax. Although the studies by \cite{Wong2019} and \cite{Wong2021}  consider the stellar interior of the WD and effects of non-conservative mass transfer, they do not investigate the evolution of systems beyond a single thermonuclear explosion. We leave a more in-depth analysis of recurrent thermonuclear SNe to future work.

If the WD reaches $M_\textrm{Ch} = 1.4 \, M_\odot$, we assume a SN~Iax occurs and examine how the dynamics of the system change due to the sudden mass ejection from the WD.  We assume no mass is stripped from the helium donor star during the explosion.  We also assume that the ejecta does not impart a recoil kick on the bound remnant.  Kicks up to several hundred ${\rm km} \, {\rm s^{-1}}$ have been found by some WD deflagration studies for certain conditions \citep{jord12a,Lach22} but not by others \citep{krom13a,Michael2014,Lach22}.  The actual recoil kick realized in nature may have a complicated dependence on initial conditions and explosion properties \citep{Lach22}; for simplicity,  we neglect it in this study.   

We use the results from \cite{hills83} for the ratio of pre- and post-SN semi-major axes and eccentricities due to a mass ejection of $\Delta M$.   Since there is ample time for circularization prior to the explosion, our initial binary systems have no eccentricity, and so we modify the \cite{hills83} result by setting the initial eccentricity $e_0 = 0$. The ratio of semi-major axes is then:
\begin{equation} \label{hills-a}
    \frac{a}{a_0} = \frac{1}{2} \left(\frac{M_0 - \Delta M}{\frac{M_0}{2} - \Delta M}\right) ,
\end{equation}
where $a$ is the post-SN semi-major axis, $a_0$ is the pre-SN semi-major axis, $M_0$ is the initial total mass of the system, and $\Delta M$ is the mass ejected from the WD due to the SN~Iax. 

The final eccentricity from \cite{hills83} is then:
\begin{equation} \label{hills-e}
    e = \left[ 1- \frac{ \left( 1- \frac{2 \Delta M}{M_0} \right)}{ \left( 1- \frac{\Delta M}{M_0} \right)^2}\right]^\frac{1}{2} .
\end{equation}

We now assume that after the SN~Iax occurs, tidal circularization happens rapidly because the helium star donor is nearly filling its Roche lobe. We calculate a new semi-major axis assuming that the orbital angular momentum is conserved as the eccentricity dissipates to zero via tides. The orbital angular momentum is given by:
\begin{equation} \label{ang-mom}
    L^2 = \frac{G (M_0 - \Delta M) \, \mu ^3}{2E} (e^2 -1) ^2 ,
\end{equation}
where $M_0 - \Delta M$ is the total mass of the system after mass ejection, $\mu$ is the reduced mass, and $E$ is the energy of the system. 

We then find the new binary separation $a_{\rm recirc}$ is given by:
\begin{equation} \label{a-recirc}
    \frac{a_{\rm recirc}}{a_0} = \frac{M_0}{M_0 - \Delta M} .
\end{equation}

Using  Equation \ref{a-recirc}, we reinitialize our MESA system with a WD mass of $1.2 \, M_\odot$, which simulates a mass ejection of $\Delta M = 0.2 \, M_\odot$, and with a new semi-major axis $a_{\rm recirc}$. We repeat this process every time the WD reaches $M_\textrm{Ch}$.   The relatively small $\Delta M = 0.2 \, M_\odot$ is on the low end of observationally inferred SN~Iax ejecta masses but matches those of low-luminosity SNe~Iax such as SN~2008ha \citep{Foley2009}.  We note that such low ejecta masses are not found by some theoretical studies \citep{jord12a,Min2014} but are produced by others \citep{Michael2014,Lach22}.  We choose this ejecta mass to maximize the possibility of recurrent SNe~Iax and leave an exploration of a range of ejecta masses to future work.

Eventually, the systems reach a natural endpoint in MESA. The most common endpoint is when the helium star ceases helium-burning, evolves into a second white dwarf, and cools. Gravitational wave radiation will cause the resulting double WD systems to merge and possibly result in a SN~Ia. The merging times of such systems are:
\begin{equation} \label{t-merge}
    t_{merge} = \frac{5c^5}{256G^3} \frac{a^4}{M_1M_2M_{tot}} .
\end{equation}

We count a SN~Ia as occurring if two conditions are met. The first is if this merging time is shorter than a Hubble time. The second is if the mass of the primary WD is greater than $\sim 0.85 \, M_\odot$ and the secondary WD is also suitably massive \citep{Dan2012, Guillochon2010, Shen17, Shen2018}, which are always the case for the systems we consider. 

Another possible endpoint is when the helium star expands into a giant and causes the mass transfer rate to increase rapidly, which MESA is unable to handle self-consistently, preventing us from simulating the full evolution of these systems. Physically, when the mass transfer rate becomes very high, the white dwarf is unable to stably accrete, causing the material to accumulate in the circumbinary environment and yielding a common envelope \citep{Iben1993, Ivanova2013}. Frictional forces from the common envelope cause the two stars to spiral towards each other. The difference in orbital energy is used to eject the envelope. Thus, for these systems in our simulation, we assume a common envelope with the mass of the unburned helium envelope of the donor star is formed and ejected, which therefore alters the binary separation. The remaining carbon core of the donor star is assumed to become a degenerate carbon-oxygen (CO) WD. Due to the assumption in our simulation that mass transfer onto the WD is stable at any accretion rate handled by MESA, we expect to see fewer common envelopes in our simulation than in reality.

We assume that the entire helium envelope is ejected and find the binding energy of that envelope:
\begin{equation}
    E_{bind} \sim - \frac{GM_{\rm He\, core}M_{\rm He\, env}}{R_{\rm He\, star}} .
\end{equation}

\begin{figure*}
  \centering
  \includegraphics[width=\textwidth]{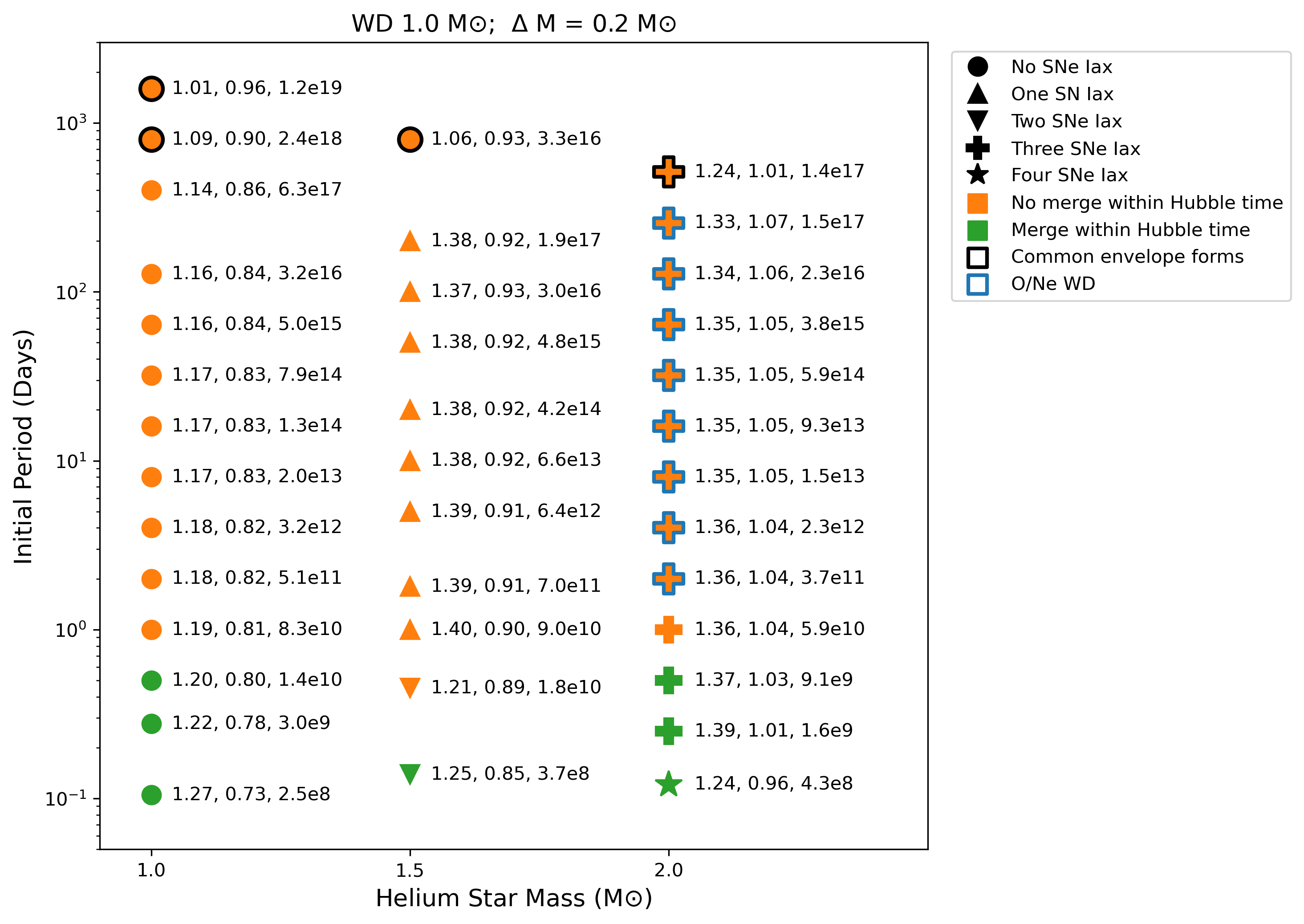}
  \caption{Summary of simulation results for various combinations of helium star mass and initial period. Fixed parameters are an initial WD mass of $1.0 \, M_\odot$ and ejected mass $\delta m = 0.2 \, M_\odot$. Text to the right of each symbol represents the final mass of the WD point mass in solar masses, the final mass of the WD that the helium star evolves into in solar masses, and the merging time of the double WD binary in years. Symbols are summarized in the legend.}
  \label{fig:Paraspace_WD1}
  \vspace{0.2in}
\end{figure*}

We then assume that in order to unbind the envelope, orbital energy is lost. Thus, we can find the relationship between the initial and final binary separations:
\begin{equation}
    \frac{E_{\rm bind}}{\alpha} = \frac{GM_{\rm He\,star}M_{\rm WD}}{a_i} - \frac{GM_{\rm He\,core}M_{\rm WD}}{a_f} ,
\end{equation}
where $\alpha$ is an efficiency parameter that we assume to be 1, and we assume the commonly used structure parameter, $\lambda$, is also 1.

After the common envelope ejection, the binary becomes a double WD system, and we calculate the expected merging time using Equation \ref{t-merge}. 

The final possible endpoint we consider is the evolution of the helium star through carbon-burning to become an oxygen/neon/magnesium (O/Ne) WD. The evolution towards an O/Ne WD is signalled by the ignition of off-center convective carbon-burning in the interior of the helium star. The calculation of such models in MESA becomes very time-intensive, so we do not further evolve these systems. However, since in each of our simulations at the stage where carbon-burning occurs, mass transfer has already decreased to negligible amounts, we assume that these carbon-burning stars in binary systems will become O/Ne WDs with the same mass.


\section{Results}
\label{sec:results}

We find that systems of recurrent SNe~Iax are plausible. In particular, in systems with greater helium star masses and smaller initial periods, the potential for several SNe~Iax increases. 

However, our estimates are optimistic. We made several approximations that increase the likelihood of multiple SNe~Iax. First, we assume only $0.2 \, M_\odot$ of material is lost from the WD during each SNe~Iax. Such a small ejecta mass may be the case for very faint SNe~Iax such as SN~2008ha \citep{Foley2009}, but typical SNe~Iax ejecta masses are likely several times larger \citep{Foley2013}.  Second, we assume that no mass is stripped from the helium donor during the SN event.   Third, we assume that all mass transferred off the helium star remains on the WD. In nature, the range of accretion rates leading to steady and stable helium-burning is only a factor of a few \citep{Brooks2016}; outside of this regime, mass is lost from the system via a combination of helium novae, optically thick winds, and binary-driven mass loss.   

Figure \ref{fig:Paraspace_WD1} demonstrates a summary of our findings. The data points represent each MESA simulation with their symbol meanings explained in the legend. The text to the right of each data point represents the final mass of the WD point mass in solar masses, the final mass of the WD that results from the evolution of the helium star in solar masses, and the merging time of the subsequent double WD binary in years. 

\subsection{Systems with $1.0 \, M_\odot$ helium star donors}

Let us examine the possible scenarios for a $1.0 \, M_\odot$ helium star. None of these systems undergo a SNe~Iax because the WD does not become massive enough to explode prior to the cessation of mass transfer and the evolution of the helium star to degeneracy. However, for systems with an initial period of $\leq$ 0.5 days, the two resulting degenerate stars will merge within a Hubble time due to the loss of orbital angular momentum by gravitational wave emission, and possibly yield a SN~Ia. Systems with an initial period $\geq$ 800 days go through a common envelope phase. Since the ejected mass is only a few hundredths of a solar mass in both cases we explored, the effect of the common envelope ejection on the merging timescale is minimal. 

\begin{figure}
  \centering
  \includegraphics[width=\columnwidth]{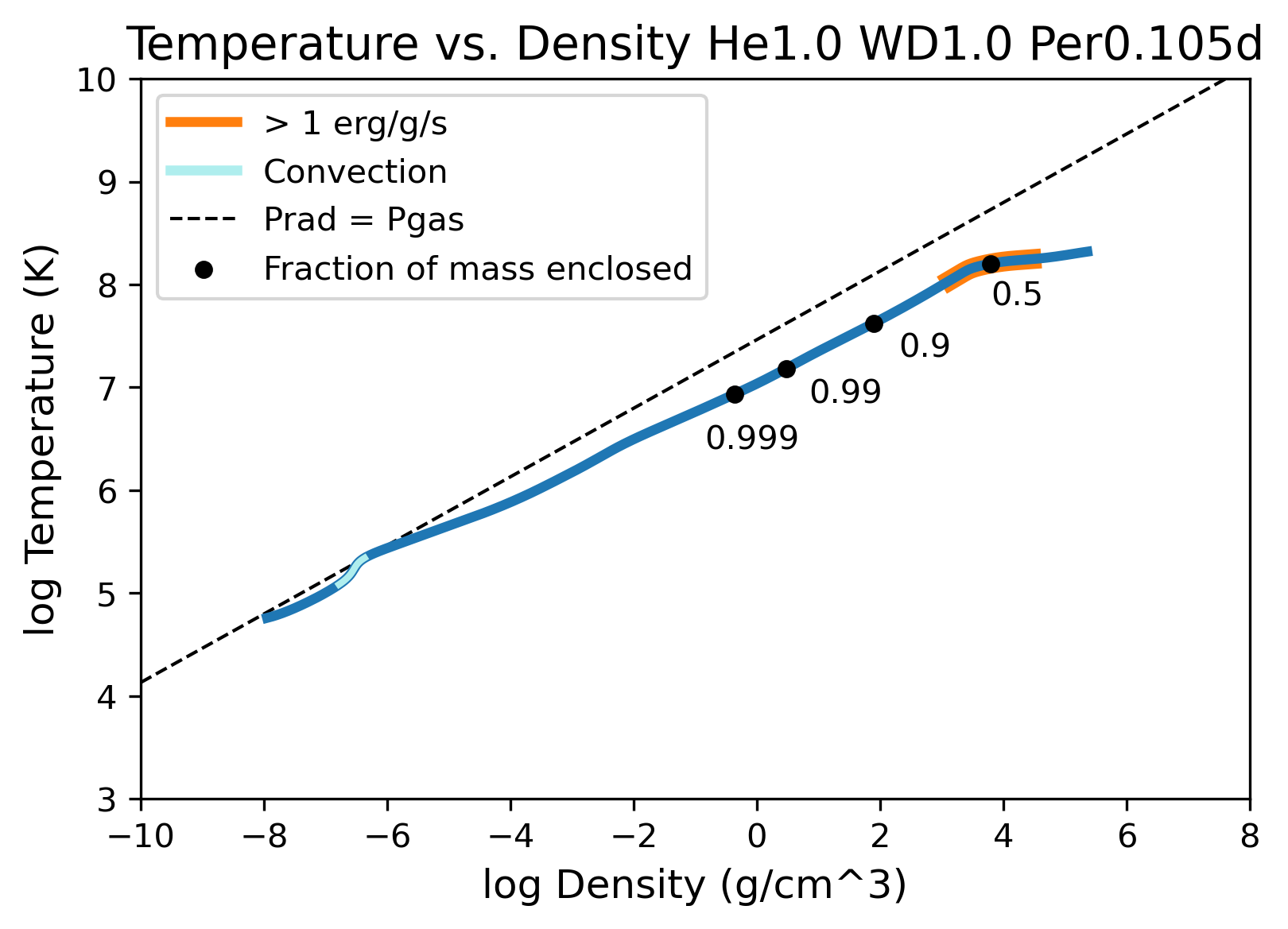}
  \caption{Temperature vs.\ density profile at the time when mass transfer rate is the highest for a system with a $1.0 \, M_\odot$ helium star with an initial period of 0.105 days. Zones where the net nuclear energy (produced from nuclear reactions minus neutrino losses) is greater than $\unit[1]{erg \, g^{-1} \, s^{-1}}$ are shown in orange. The dashed line shows conditions for which the radiation pressure equals the ideal gas pressure, assuming a chemical composition of fully ionized helium. Labelled black dots represent mass fractions contained within a certain zone.}
  \label{fig:Hertzsprung}
\end{figure}

\begin{figure}
  \centering
  \includegraphics[width=\columnwidth]{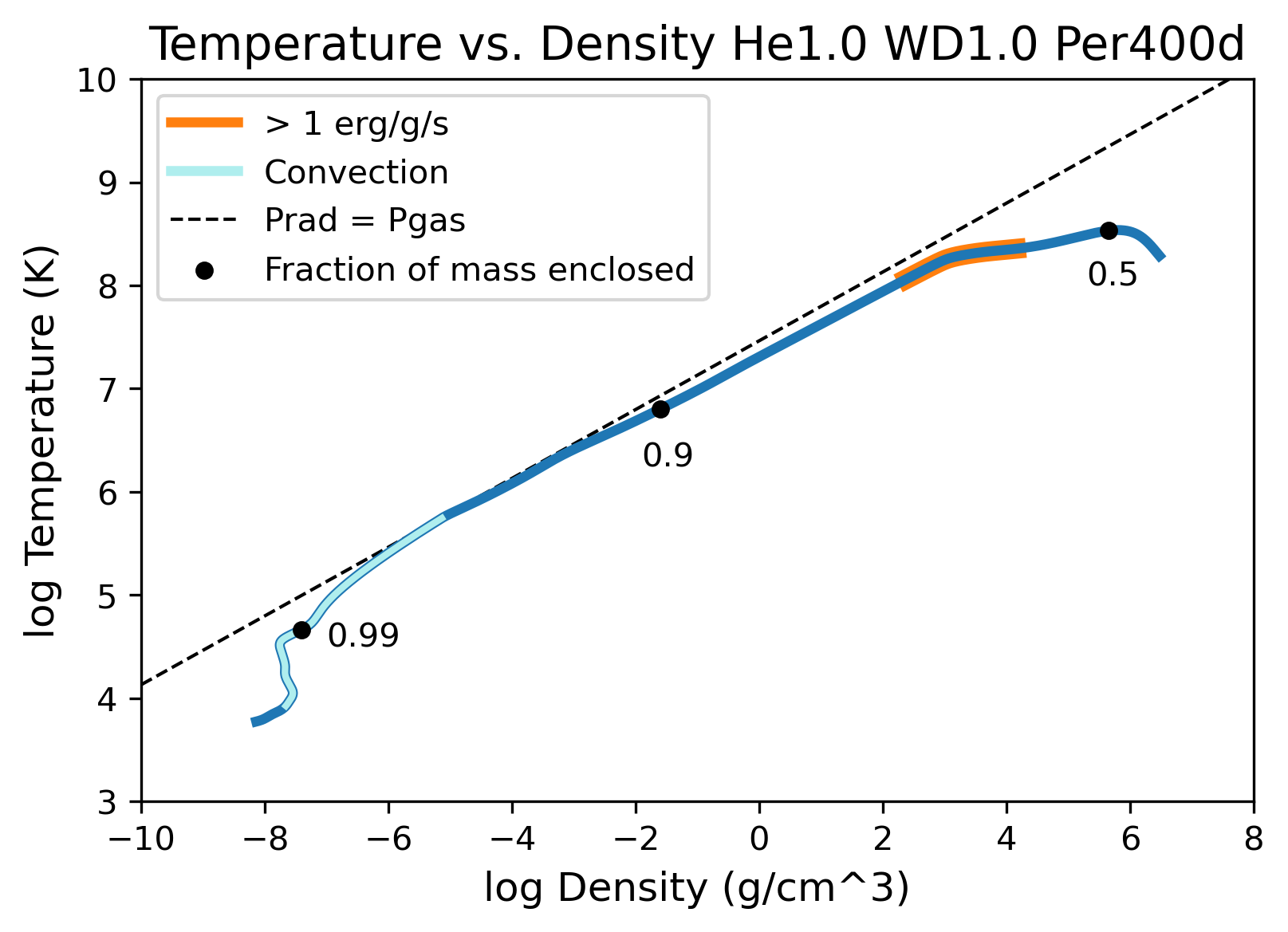}
  \caption{Temperature vs.\ density profile at the time when mass transfer rate is the highest for a system with a $1.0 \, M_\odot$ helium star with an initial period of 400 days. In this profile, labelled mass fractions show a dense core and massive envelope, which demonstrates that the star is in the giant phase.}
  \label{fig:giant}
\end{figure}

For a $1.0 \, M_\odot$ helium star, when the initial period is small, such as in the case of an initial period of 0.105 days, mass transfer begins as soon as the helium star begins to expand. Thus, the bulk of mass transfer occurs along the Hertzsprung gap before the star becomes a giant. This is evident in the temperature vs.\ density profile of the star at the time when the mass transfer rate is highest, which is shown in Figure \ref{fig:Hertzsprung}. In this profile, we see that the star is no longer on the main sequence because the burning region is outside the core. However, the star is not yet a giant because much of the mass is confined within a relatively compact inner region of the star. The density at an enclosed mass fraction of 0.99 is $3~\text{g/cm}^3$. In contrast, Figure \ref{fig:giant} demonstrates the temperature vs.\ density profile of a $1.0 \, M_\odot$ helium star that is a giant at the time when the mass transfer rate is highest. This system began with a much larger initial period of 400 days, and therefore requires the star to expand significantly before it fills its Roche Lobe. The temperature vs.\ density profile shows a more massive low density envelope. The density of this star at a mass fraction of 0.99 is $4 \times 10^{-8}~\text{g/cm}^3$. 

\begin{figure}
  \centering
  \includegraphics[width=\columnwidth]{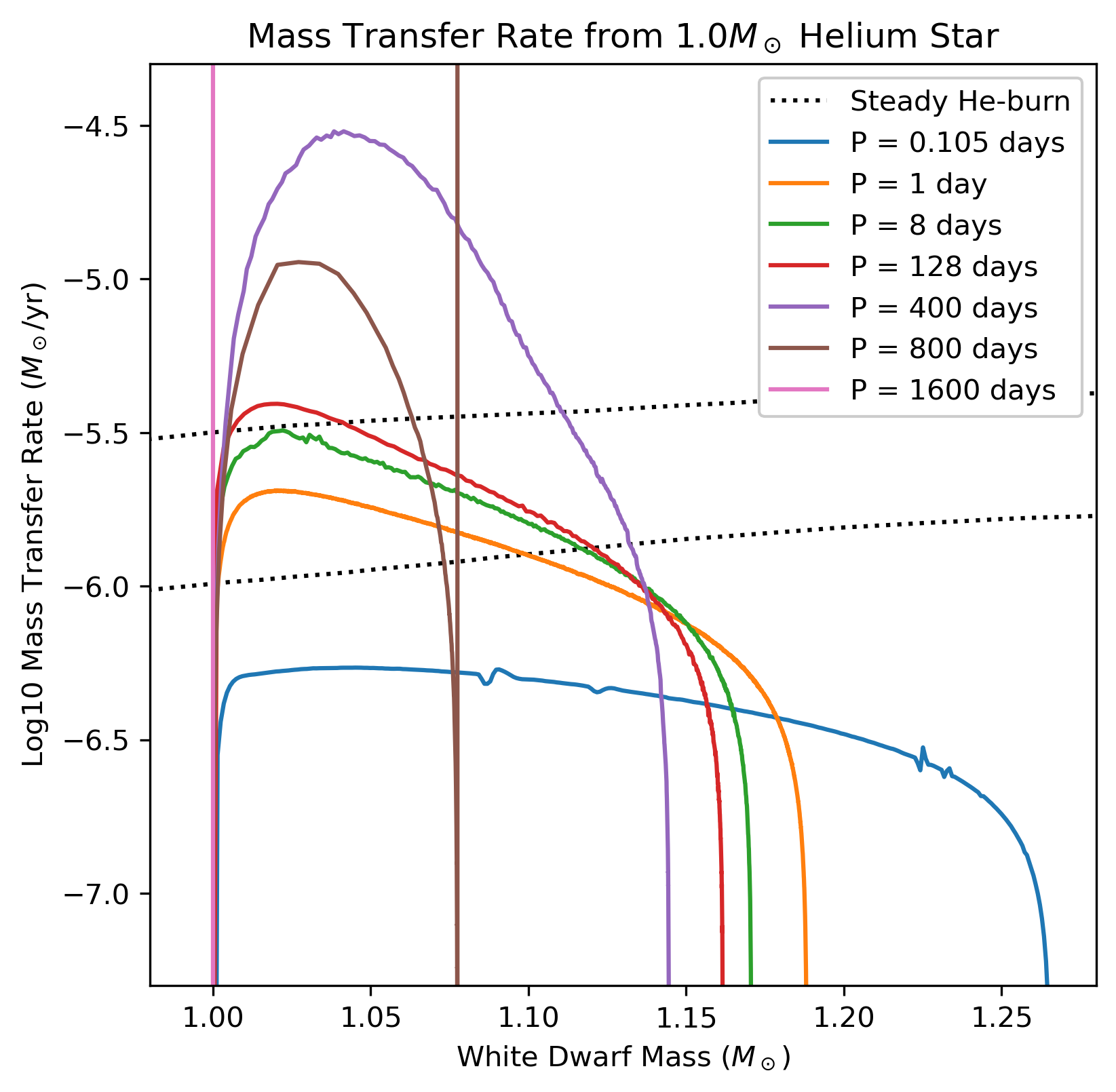}
  \caption{Mass transfer rates for systems with an initial helium star mass of $1.0 \, M_\odot$ and differing initial periods. Dotted lines are the upper and lower bounds of accretion that lead to steady helium-burning on the WD's surface according to \cite{Brooks2016}.}
  \label{fig:mdot_1p0}
\end{figure}

Figure \ref{fig:mdot_1p0} demonstrates the mass transfer rate vs.\ WD mass for several systems with an initial $1.0 \, M_\odot$ helium star. We see that none of the WDs reach the Chandrasekhar-mass. The black dotted lines represent the upper and lower boundaries of accretion required for steady and stable helium-burning on the WD according to Figure 3 from \cite{Brooks2016}. All of the systems shown in Figure \ref{fig:mdot_1p0} transfer a significant amount of mass outside these boundaries, which demonstrates the limitations of our assumption that all mass is stably transferred onto the WD point mass. The mass transfer rates run away above rates of a few times $10^{-5} M_\odot$/yr for systems with an initial period $\geq$ 800 days. 

\subsection{Systems with $1.5 \, M_\odot$ helium star donors}

Next, we examine systems with $1.5 \, M_\odot$ helium stars. Systems with an initial period of approximately $\leq$ 0.45 days can go through two sequential SNe~Iax. Within this group, systems with an initial period of approximately $\leq$ 0.139 days can also merge within a Hubble time, which allows for the potential for a SN~Ia following the previous two SNe~Iax. For systems with periods between 1 and 400 days, we only find potential for one SN~Iax based on the previously described constraints. Notably, for systems with periods of 800 days, zero SNe are expected. Additionally, a common envelope phase occurs and ejects approximately $0.5 \, M_\odot$ of material, or approximately 34\% of the helium star's original mass. This brings the system into closer orbit and reduces the merging time slightly. We also expect the common envelope phase will likely occur for systems with initial periods $\geq$ 800 days. 

\begin{figure}
  \centering
  \includegraphics[width=\columnwidth]{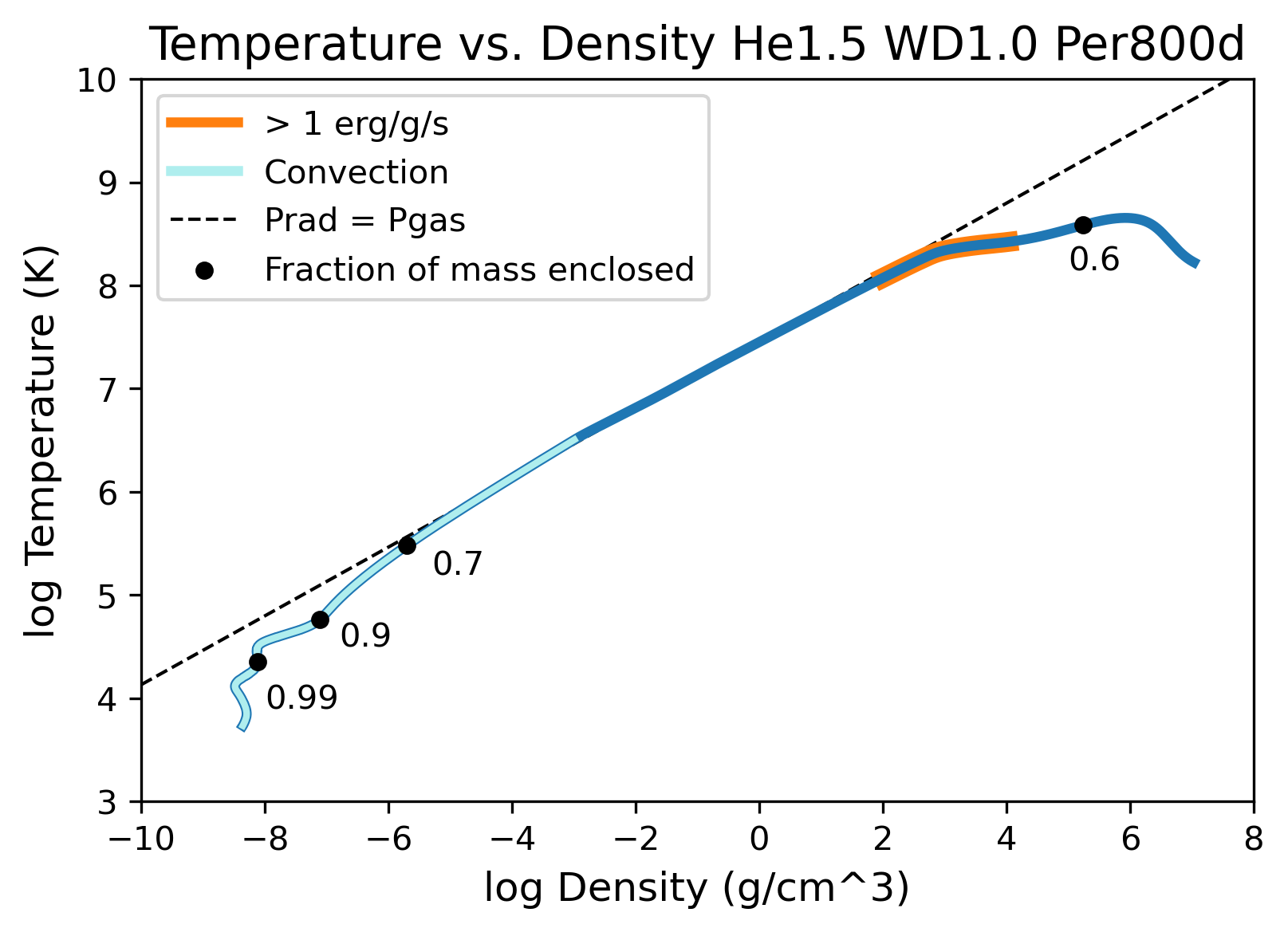}
  \caption{Temperature vs.\ density profile at the time when mass transfer rate is the highest for a system with a $1.5 \, M_\odot$ helium star with an initial period of 800 days. The mass transfer rate for this system becomes high enough to enter a common envelope phase. In this profile, labelled mass fractions show a dense core and massive envelope, which demonstrates that the star is in the giant phase. In contrast to Figure \ref{fig:giant}, a greater fraction of the mass resides in this low density envelope, which is ejected as a common envelope.}
  \label{fig:common_env}
\end{figure}

Figure \ref{fig:common_env} demonstrates the temperature vs.\ density profile for the system with an initial helium star mass of $1.5 \, M_\odot$ and an initial period of 800 days. This profile corresponds to the time when the mass transfer rate is highest, which is directly prior to the end of MESA's ability to calculate the remaining binary evolution. We notice that at least 60\% of the star's mass is in a compact region at the core of the star at a density $\geq 8 \times 10^5~\text{g/cm}^3$ and a massive envelope contains at least 30\% of the mass at a density $\leq   10^{-5}~\text{g/cm}^3$. This convective envelope contributes to runaway mass transfer and the formation of a common envelope \citep{Ge2010}. 

\begin{figure}
  \centering
  \includegraphics[width=\columnwidth]{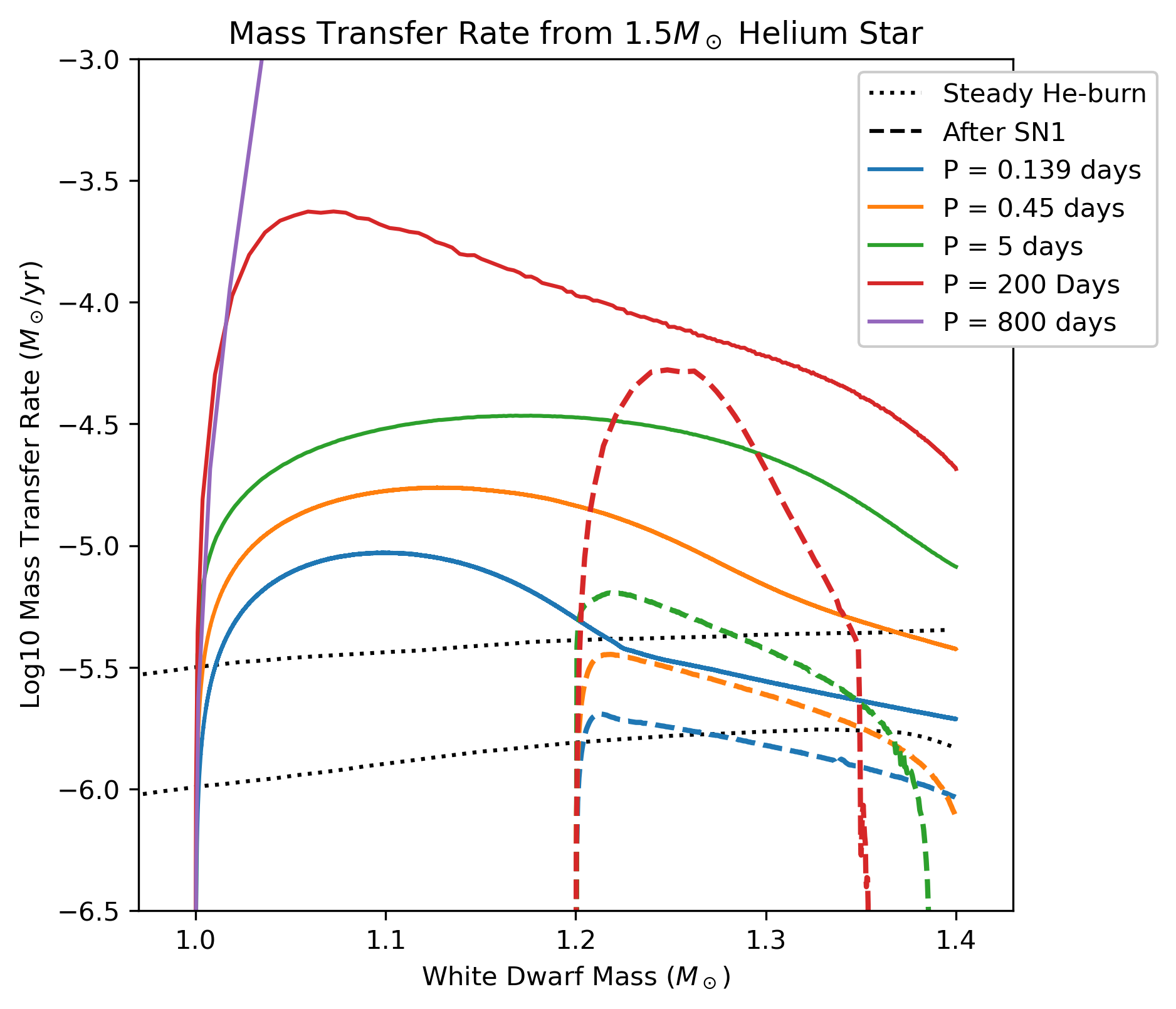}
  \caption{Mass transfer rates for systems with helium star mass of $1.5 \, M_\odot$ and differing initial periods. The mass transfer evolution prior to the first and second SN~Iax are shown as solid and dashed lines, respectively, but any further evolution is neglected for visual simplicity.}
  \label{fig:mdot_1p5}
\end{figure}

Figure \ref{fig:mdot_1p5} demonstrates the mass transfer rate vs.\ WD mass for several systems with a $1.5 \, M_\odot$ helium star. The solid lines represent the mass transfer prior to the first SN and the dashed lines represent mass transfer prior to the second SN. If a second SN occurs, all further mass transfer is neglected for visual simplicity. This figure demonstrates that most mass transfer for these systems occurs above the rate at which transfer is stable. The mass transfer rate for the system with an initial period of 800 days runs away above a few times $10^{-4} M_\odot$/yr.

\subsection{Systems with $2.0 \, M_\odot$ helium star donors}

\begin{figure}
  \centering
  \includegraphics[width=\columnwidth]{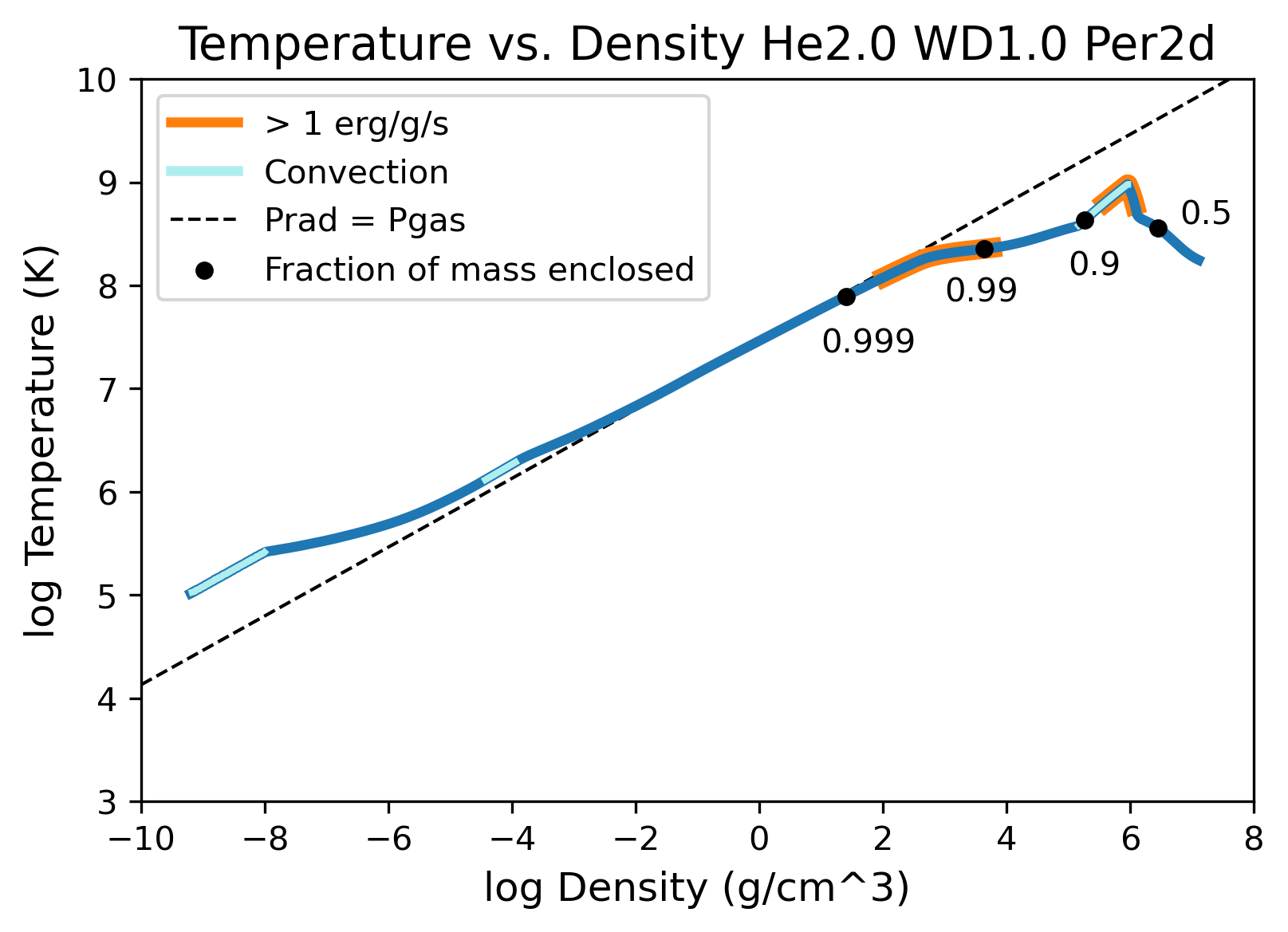}
  \caption{Temperature vs.\ density profile for a system with a $2.0 \, M_\odot$ helium star with an initial period of 2 days. Outer and inner burning locations (shown in orange) correspond to helium-burning and carbon-burning regions, respectively.}
  \label{fig:ONe}
\end{figure}

\begin{figure}
  \centering
  \includegraphics[width=\columnwidth]{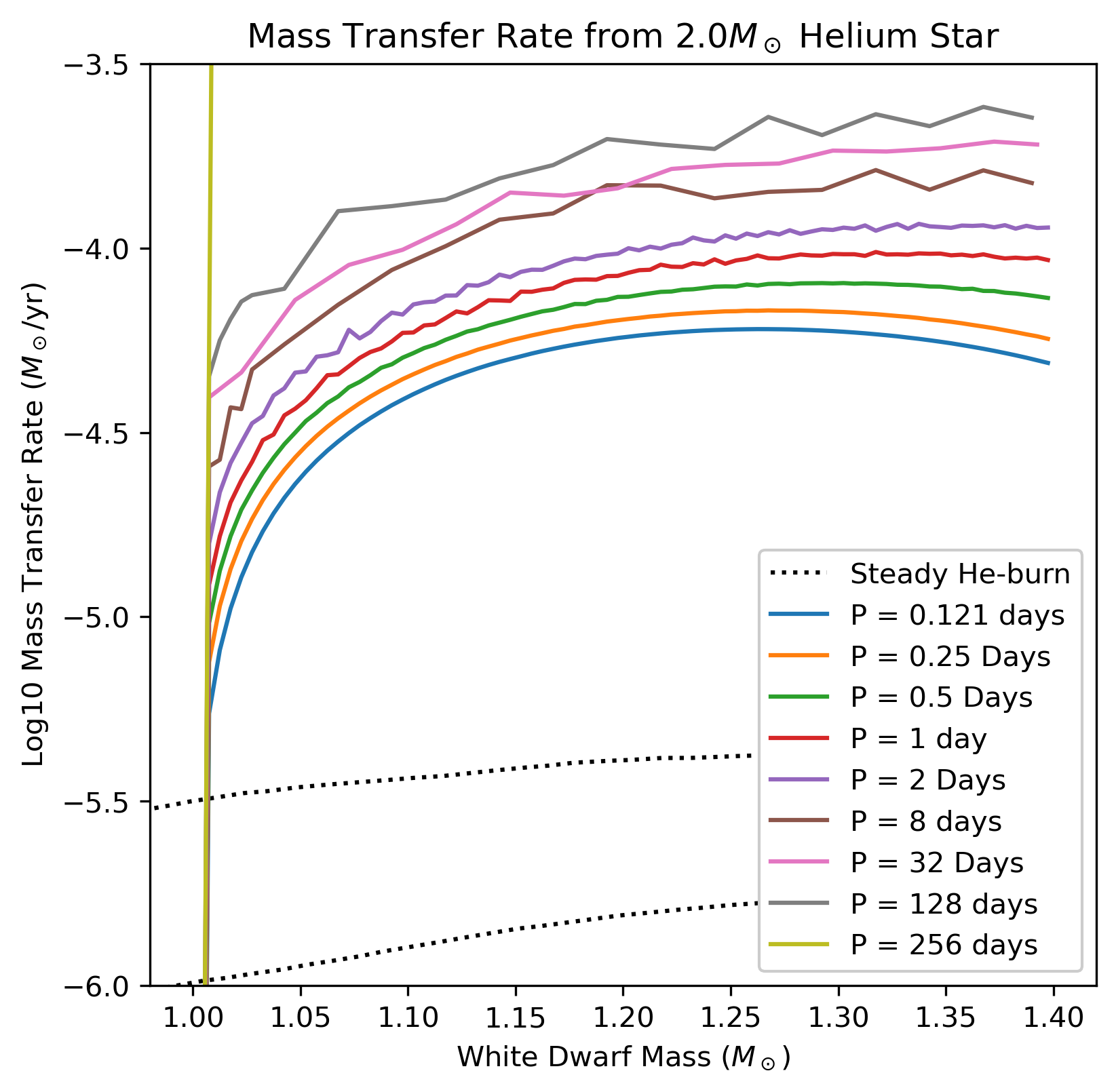}
  \caption{Mass transfer rates for systems with helium star mass of $2.0 \, M_\odot$ and differing initial periods. The mass transfer evolution beyond the first SN~Iax is neglected for visual simplicity.}
  \label{fig:mdot_2p0}
\end{figure}

Finally, consider systems with $2.0 \, M_\odot$ helium stars. Notably, for the entire range of initial periods that we explored, the systems undergo at least three Type Iax SNe. Systems with an initial period of 0.121 days have the potential for four SNe~Iax, and systems with an initial period $\leq$ 0.5 days have the potential to merge within a Hubble time. Systems with an initial period of 512 days will undergo a common envelope phase and eject $0.15 \, M_\odot$ of material. Additionally, helium stars that will eventually evolve into WDs with a mass $\geq 1.04 \,  M_\odot$ ignite core carbon-burning. The inward propagation of the carbon-burning wave by diffusion results in very small simulation timesteps, and therefore we halt these calculations before their  natural endpoints. However, we can assume that carbon-burning will cause the star to become an oxygen/neon/magnesium (O/Ne) WD. The temperature vs.\ density profile for a star becoming an O/Ne WD is shown in Figure \ref{fig:ONe}. The outer and inner burning regions correspond to helium-burning and convective carbon-burning regions, respectively. Much of the mass of this star is at a high density. The density at an enclosed mass fraction of 0.99 is $4 \times 10^3~\text{g/cm}^3$. 

Figure \ref{fig:mdot_2p0} demonstrates the mass transfer rate vs.\ WD mass for several systems with a $2.0 \, M_\odot$ helium star. Only the mass transfer prior to the first SN is shown for visual simplicity. From this figure, we can determine that none of the mass transfer from $2.0 \, M_\odot$ helium stars occurs in the stable region prior to the first SN, which highlights the extreme optimism of our assumptions. In reality, since mass transfer occurs above the stable boundary, we expect the formation of an optically thick wind or a  common envelope, which we did not account for in systems with a period $< 512$ days. 


\section{Conclusions}
\label{sec:conc}

In this investigation, we used MESA to explore the potential for recurrent thermonuclear SNe from binary systems containing a helium star donor and WD accretor. Using our optimistic assumptions, we found that the possibility for several SNe~Iax exists in systems with high-mass helium stars and small initial periods. We also find that systems with small initial periods are likely to produce SNe~Ia by forming double WD systems and merging within a Hubble time, suggesting the possibility for both single- and double-degenerate thermonuclear SNe produced by the same system. 

However, we also find that the systems most likely to create recurrent SNe~Iax have the highest degree of error in assumptions in our simulation. Each of these systems transfers mass at rates above the WD stable accretion rate set by steady and stable helium-burning on the surface of the WD. Thus, our assumption of conservative mass transfer onto the WD is very optimistic. Likely, these systems would undergo a common envelope phase and eject much of the unaccreted material, which would greatly decrease the chances for repeated SNe~Iax. 

Our simulation is also optimistic due to the treatment of the WD as a point mass, and resulting inability to determine whether carbon ignition begins centrally or off-center. In cases of off-center carbon ignition, the WD burns non-explosively and is converted into an O/Ne WD \citep{Brooks2016} instead of leading to a SN.

The possibility that binaries can host multiple thermonuclear SNe has obvious implications for predictions of SN rates and galactic chemical enrichment. This possibility also suggests polluted helium stars in WD binaries may exist and may be observable throughout the galaxy.  Furthermore, thermonuclear ash from previous SNe that remains inside a surviving WD that subsequently undergoes another SN may alter the resulting nucleosynthesis in interesting ways.  However, given the small parameter space that leads to multiple SNe (restricted to short orbital periods and  massive helium star donors) and the overly optimistic assumptions employed in this work, it appears unlikely that systems with multiple SNe play a significant role in nature. Still, given the exciting ramifications that are possible, we recommend that future studies be performed with more accurate physics regarding the stability and retention of accreted helium.


\acknowledgments

This project was initiated with Saurabh Jha and Todd Thompson during a Scialog Time Domain Astrophysics workshop.  We further thank Todd Thompson for helpful comments that improved this manuscript.  Support for this work was provided by NASA/ESA Hubble Space Telescope program \#15918 and by the National Science Foundation under Grant No.\ NSF PHY-1748958.


\software{\texttt{matplotlib} \citep{hunt07a}, \texttt{MESA} \citep{paxt11,paxt13,paxt15a,paxt18a,paxt19a}}


\appendix
\label{appendix}

The following inlist\_project was used to create helium star models in MESA: \\

{\ttfamily
\noindent \&star\_job \\
 \\
\indent ! do not begin with a pre-main sequence model \\
\indent create\_pre\_main\_sequence\_model = .false. \\
 \\
\indent ! save a model once helium star is created \\
\indent save\_model\_when\_terminate = .false. \\
\indent load\_saved\_model = .false. \\
\indent save\_model\_number = 10 \\
\indent save\_model\_filename = 'He1.0.mod' \\
 \\
\indent ! display on-screen plots \\
\indent pgstar\_flag = .true. \\
 \\
\indent ! change chemical composition of the star \\
\indent relax\_initial\_to\_xaccrete = .true. \\
 \\
\noindent / !end of star\_job namelist \\

\noindent \&controls \\
 \\
\indent ! set initial mass \\
\indent initial\_mass = 1.0 ! in Msun units \\
 \\
\indent ! options for energy conservation (see MESA V, Section 3) \\
\indent use\_dedt\_form\_of\_energy\_eqn = .true. \\
 \\
\indent ! set chemical composition of the star \\
\indent accrete\_given\_mass\_fractions = .true. \\
\indent num\_accretion\_species = 2 \\
\indent accretion\_species\_id(1) = 'he4' \\
\indent accretion\_species\_xa(1) = 0.98 \\
\indent accretion\_species\_id(2) = 'n14' \\
\indent accretion\_species\_xa(2) = 0.02 \\
 \\
\noindent / ! end of controls namelist \\
}

When our binary systems encountered difficulties in MESA, a few settings were toggled to allow the evolution to occur more smoothly. These included the allowed variation in stellar structure between models (\texttt{varcontrol\_target}), the method of numerical calculation (\texttt{min\_mdot\_for\_implicit}), and the efficiency of energy transport (\texttt{okay\_to\_reduce\_gradT\_excess}). When some systems encountered unstable periods, we briefly adjusted the convective mixing length (\texttt{mixing\_length\_alpha}) in order to allow the system to progress. 

The following inlist\_project is an example of how we evolved our binaries: \\

{\ttfamily 
\noindent \&binary\_job \\
 \\
\indent inlist\_names(1) = 'inlist1' \\
\indent inlist\_names(2) = 'inlist2' \\
 \\
\indent evolve\_both\_stars = .false. \\
 \\
\noindent / ! end of binary\_job namelist \\

\noindent \&binary\_controls \\
 \\
\indent terminate\_if\_initial\_overflow = .false. \\
 \\
\indent m2 = 1.0d0 ! point mass in Msun; reduced by 0.2 in each sequential run to account for mass loss from SN \\
\indent initial\_period\_in\_days = 1 \\
\indent initial\_separation\_in\_Rsuns = 100 ! use initial separation calculated after SNe, rather than initial period \\
 \\
\indent ! angular momentum calculations \\
\indent do\_jdot\_mb = .false. \\
\indent do\_jdot\_gr = .true. \\
\indent do\_jdot\_ml = .false. \\
\indent do\_jdot\_ls = .false. \\
 \\
\indent ! set whether calculation is implicit or explicit \\
\indent min\_mdot\_for\_implicit = 1d99 ! default is 1d-16; when mdot in Msun/secyer > min\_mdot\_for\_implicit, then calculation is explicit \\
 \\
\noindent / ! end of binary\_controls namelist \\
}

The following inlist is an example of how we evolved our helium stars in the binary systems: \\

{\ttfamily
\noindent \&star\_job \\
 \\
\indent ! begin from a previously created helium model \\
\indent load\_saved\_model = .true. \\
\indent saved\_model\_name = 'He1.0.mod' \\
 \\
\indent ! if the helium star reaches a mass such that the point mass is 1.4 Msun, then terminate the run and save the helium star model \\
\indent save\_model\_when\_terminate = .true. \\
\indent required\_termination\_code\_string = 'star\_mass\_min\_limit' \\
\indent save\_model\_filename = 'He1.0\_WD1.0\_per1\_SN1.mod' \\
 \\
\noindent / ! end of star\_job namelist \\

\noindent \&controls \\
 \\
\indent ! minimum mass for helium star such that point mass reaches maximum of 1.4 Msun \\
\indent star\_mass\_min\_limit = 0.6d0 \\
 \\
\indent okay\_to\_reduce\_gradT\_excess = .false. ! turned to .true.\ for the entire run if energy transport is causing run to evolve too slowly\\
 \\
\indent varcontrol\_target = 1d-4 ! increased to 3d-4 if varcontrol is the limiter of evolution between steps\\
 \\
\indent mixing\_length\_alpha = 2 ! increased to 4 or 8 for brief periods to move past structural instability of helium star\\
 \\
\noindent / ! end of controls namelist \\
}




\begin{thebibliography}{}
\expandafter\ifx\csname natexlab\endcsname\relax\def\natexlab#1{#1}\fi
\providecommand{\url}[1]{\href{#1}{#1}}
\providecommand{\dodoi}[1]{doi:~\href{http://doi.org/#1}{\nolinkurl{#1}}}
\providecommand{\doeprint}[1]{\href{http://ascl.net/#1}{\nolinkurl{http://ascl.net/#1}}}
\providecommand{\doarXiv}[1]{\href{https://arxiv.org/abs/#1}{\nolinkurl{https://arxiv.org/abs/#1}}}

\bibitem[{{Bauer} {et~al.}(2019){Bauer}, {White}, \& {Bildsten}}]{Bauer19}
{Bauer}, E.~B., {White}, C.~J., \& {Bildsten}, L. 2019, \apj, 887, 68,
  \dodoi{10.3847/1538-4357/ab4ea4}

\bibitem[{{Bobrick} {et~al.}(2022){Bobrick}, {Zenati}, {Perets}, {Davies}, \&
  {Church}}]{bobr22a}
{Bobrick}, A., {Zenati}, Y., {Perets}, H.~B., {Davies}, M.~B., \& {Church}, R.
  2022, \mnras, 510, 3758,
  \dodoi{10.1093/mnras/stab357410.48550/arXiv.2104.03415}

\bibitem[{{Brooks} {et~al.}(2016){Brooks}, {Bildsten}, {Schwab}, \&
  {Paxton}}]{Brooks2016}
{Brooks}, J., {Bildsten}, L., {Schwab}, J., \& {Paxton}, B. 2016, \apj, 821,
  28, \dodoi{10.3847/0004-637X/821/1/28}

\bibitem[{{Brooks} {et~al.}(2017){Brooks}, {Schwab}, {Bildsten}, {Quataert}, \&
  {Paxton}}]{broo17a}
{Brooks}, J., {Schwab}, J., {Bildsten}, L., {Quataert}, E., \& {Paxton}, B.
  2017, \apjl, 834, L9

\bibitem[{{Dan} {et~al.}(2012){Dan}, {Rosswog}, {Guillochon}, \&
  {Ramirez-Ruiz}}]{Dan2012}
{Dan}, M., {Rosswog}, S., {Guillochon}, J., \& {Ramirez-Ruiz}, E. 2012, \mnras,
  422, 2417, \dodoi{10.1111/j.1365-2966.2012.20794.x}

\bibitem[{{Denissenkov} {et~al.}(2013){Denissenkov}, {Herwig}, {Truran}, \&
  {Paxton}}]{deni13b}
{Denissenkov}, P.~A., {Herwig}, F., {Truran}, J.~W., \& {Paxton}, B. 2013,
  \apj, 772, 37

\bibitem[{{Fink} {et~al.}(2014){Fink}, {Kromer}, {Seitenzahl},
  {Ciaraldi-Schoolmann}, {R{\"o}pke}, {Sim}, {Pakmor}, {Ruiter}, \&
  {Hillebrandt}}]{Michael2014}
{Fink}, M., {Kromer}, M., {Seitenzahl}, I.~R., {et~al.} 2014, \mnras, 438,
  1762, \dodoi{10.1093/mnras/stt2315}

\bibitem[{{Foley} {et~al.}(2016){Foley}, {Jha}, {Pan}, {Zheng}, {Bildsten},
  {Filippenko}, \& {Kasen}}]{Foley2016}
{Foley}, R.~J., {Jha}, S.~W., {Pan}, Y.-C., {et~al.} 2016, \mnras, 461, 433,
  \dodoi{10.1093/mnras/stw1320}

\bibitem[{{Foley} {et~al.}(2009){Foley}, {Chornock}, {Filippenko},
  {Ganeshalingam}, {Kirshner}, {Li}, {Cenko}, {Challis}, {Friedman}, {Modjaz},
  {Silverman}, \& {Wood-Vasey}}]{Foley2009}
{Foley}, R.~J., {Chornock}, R., {Filippenko}, A.~V., {et~al.} 2009, \aj, 138,
  376, \dodoi{10.1088/0004-6256/138/2/376}

\bibitem[{{Foley} {et~al.}(2010){Foley}, {Rest}, {Stritzinger}, {Pignata},
  {Anderson}, {Hamuy}, {Morrell}, {Phillips}, \& {Salgado}}]{Foley10}
{Foley}, R.~J., {Rest}, A., {Stritzinger}, M., {et~al.} 2010, \aj, 140, 1321,
  \dodoi{10.1088/0004-6256/140/5/1321}

\bibitem[{{Foley} {et~al.}(2013){Foley}, {Challis}, {Chornock},
  {Ganeshalingam}, {Li}, {Marion}, {Morrell}, {Pignata}, {Stritzinger},
  {Silverman}, {Wang}, {Anderson}, {Filippenko}, {Freedman}, {Hamuy}, {Jha},
  {Kirshner}, {McCully}, {Persson}, {Phillips}, {Reichart}, \&
  {Soderberg}}]{Foley2013}
{Foley}, R.~J., {Challis}, P.~J., {Chornock}, R., {et~al.} 2013, \apj, 767, 57,
  \dodoi{10.1088/0004-637X/767/1/57}

\bibitem[{{Ge} {et~al.}(2010){Ge}, {Hjellming}, {Webbink}, {Chen}, \&
  {Han}}]{Ge2010}
{Ge}, H., {Hjellming}, M.~S., {Webbink}, R.~F., {Chen}, X., \& {Han}, Z. 2010,
  \apj, 717, 724, \dodoi{10.1088/0004-637X/717/2/724}

\bibitem[{{Guillochon} {et~al.}(2010){Guillochon}, {Dan}, {Ramirez-Ruiz}, \&
  {Rosswog}}]{Guillochon2010}
{Guillochon}, J., {Dan}, M., {Ramirez-Ruiz}, E., \& {Rosswog}, S. 2010, \apjl,
  709, L64, \dodoi{10.1088/2041-8205/709/1/L64}

\bibitem[{{Hills}(1983)}]{hills83}
{Hills}, J.~G. 1983, \apj, 267, 322, \dodoi{10.1086/160871}

\bibitem[{Hunter(2007)}]{hunt07a}
Hunter, J.~D. 2007, Computing in Science \& Engineering, 9, 90,
  \dodoi{10.1109/MCSE.2007.55}

\bibitem[{{Iben} \& {Livio}(1993)}]{Iben1993}
{Iben}, Icko, J., \& {Livio}, M. 1993, \pasp, 105, 1373, \dodoi{10.1086/133321}

\bibitem[{{Iben} \& {Tutukov}(1994)}]{IbenTutukov1994}
{Iben}, Icko, J., \& {Tutukov}, A.~V. 1994, \apj, 431, 264,
  \dodoi{10.1086/174484}

\bibitem[{{Ivanova} {et~al.}(2013){Ivanova}, {Justham}, {Chen}, {De Marco},
  {Fryer}, {Gaburov}, {Ge}, {Glebbeek}, {Han}, {Li}, {Lu}, {Marsh},
  {Podsiadlowski}, {Potter}, {Soker}, {Taam}, {Tauris}, {van den Heuvel}, \&
  {Webbink}}]{Ivanova2013}
{Ivanova}, N., {Justham}, S., {Chen}, X., {et~al.} 2013, \aapr, 21, 59,
  \dodoi{10.1007/s00159-013-0059-2}

\bibitem[{{Jha}(2017)}]{Jha2017}
{Jha}, S.~W. 2017, in Handbook of Supernovae, ed. A.~W. {Alsabti} \&
  P.~{Murdin}, 375, \dodoi{10.1007/978-3-319-21846-5\_42}

\bibitem[{{Jordan} {et~al.}(2012){Jordan}, {Perets}, {Fisher}, \& {van
  Rossum}}]{jord12a}
{Jordan}, IV, G.~C., {Perets}, H.~B., {Fisher}, R.~T., \& {van Rossum}, D.~R.
  2012, \apjl, 761, L23

\bibitem[{{Kashyap} {et~al.}(2018){Kashyap}, {Haque}, {Lor{\'e}n-Aguilar},
  {Garc{\'\i}a-Berro}, \& {Fisher}}]{kash18a}
{Kashyap}, R., {Haque}, T., {Lor{\'e}n-Aguilar}, P., {Garc{\'\i}a-Berro}, E.,
  \& {Fisher}, R. 2018, \apj, 869, 140,
  \dodoi{10.3847/1538-4357/aaedb710.48550/arXiv.1811.00013}

\bibitem[{{Kromer} {et~al.}(2013){Kromer}, {Fink}, {Stanishev}, {Taubenberger},
  {Ciaraldi-Schoolman}, {Pakmor}, {R{\"o}pke}, {Ruiter}, {Seitenzahl}, {Sim},
  {Blanc}, {Elias-Rosa}, \& {Hillebrandt}}]{krom13a}
{Kromer}, M., {Fink}, M., {Stanishev}, V., {et~al.} 2013, \mnras, 429, 2287

\bibitem[{{Kromer} {et~al.}(2015){Kromer}, {Ohlmann}, {Pakmor}, {Ruiter},
  {Hillebrandt}, {Marquardt}, {R{\"o}pke}, {Seitenzahl}, {Sim}, \&
  {Taubenberger}}]{krom15a}
{Kromer}, M., {Ohlmann}, S.~T., {Pakmor}, R., {et~al.} 2015, \mnras, 450, 3045

\bibitem[{{Lach} {et~al.}(2022){Lach}, {Callan}, {Bubeck}, {R{\"o}pke}, {Sim},
  {Schrauth}, {Ohlmann}, \& {Kromer}}]{Lach22}
{Lach}, F., {Callan}, F.~P., {Bubeck}, D., {et~al.} 2022, \aap, 658, A179,
  \dodoi{10.1051/0004-6361/202141453}

\bibitem[{{Lecoanet} {et~al.}(2016){Lecoanet}, {Schwab}, {Quataert},
  {Bildsten}, {Timmes}, {Burns}, {Vasil}, {Oishi}, \& {Brown}}]{leco16a}
{Lecoanet}, D., {Schwab}, J., {Quataert}, E., {et~al.} 2016, \apj, 832, 71

\bibitem[{{Li} {et~al.}(2003){Li}, {Filippenko}, {Chornock}, {Berger},
  {Berlind}, {Calkins}, {Challis}, {Fassnacht}, {Jha}, {Kirshner}, {Matheson},
  {Sargent}, {Simcoe}, {Smith}, \& {Squires}}]{Li03}
{Li}, W., {Filippenko}, A.~V., {Chornock}, R., {et~al.} 2003, \pasp, 115, 453,
  \dodoi{10.1086/374200}

\bibitem[{{Liu} {et~al.}(2015){Liu}, {Stancliffe}, {Abate}, \&
  {Wang}}]{Liu2015}
{Liu}, Z.-W., {Stancliffe}, R.~J., {Abate}, C., \& {Wang}, B. 2015, \apj, 808,
  138, \dodoi{10.1088/0004-637X/808/2/138}

\bibitem[{{Liu} {et~al.}(2013){Liu}, {Pakmor}, {Seitenzahl}, {Hillebrandt},
  {Kromer}, {R{\"o}pke}, {Edelmann}, {Taubenberger}, {Maeda}, {Wang}, \&
  {Han}}]{Liu13}
{Liu}, Z.-W., {Pakmor}, R., {Seitenzahl}, I.~R., {et~al.} 2013, \apj, 774, 37,
  \dodoi{10.1088/0004-637X/774/1/37}

\bibitem[{{Long} {et~al.}(2014){Long}, {Jordan}, {van Rossum}, {Diemer},
  {Graziani}, {Kessler}, {Meyer}, {Rich}, \& {Lamb}}]{Min2014}
{Long}, M., {Jordan}, George~C., I., {van Rossum}, D.~R., {et~al.} 2014, \apj,
  789, 103, \dodoi{10.1088/0004-637X/789/2/103}

\bibitem[{{McCully} {et~al.}(2014){McCully}, {Jha}, {Foley}, {Bildsten},
  {Fong}, {Kirshner}, {Marion}, {Riess}, \& {Stritzinger}}]{McCully2014}
{McCully}, C., {Jha}, S.~W., {Foley}, R.~J., {et~al.} 2014, \nat, 512, 54,
  \dodoi{10.1038/nature13615}

\bibitem[{{Meng} \& {Podsiadlowski}(2014)}]{meng14a}
{Meng}, X., \& {Podsiadlowski}, P. 2014, \apjl, 789, L45

\bibitem[{{Pan} {et~al.}(2012){Pan}, {Ricker}, \& {Taam}}]{Pan12}
{Pan}, K.-C., {Ricker}, P.~M., \& {Taam}, R.~E. 2012, \apj, 750, 151,
  \dodoi{10.1088/0004-637X/750/2/151}

\bibitem[{{Pan} {et~al.}(2013){Pan}, {Ricker}, \& {Taam}}]{Pan13}
---. 2013, \apj, 773, 49, \dodoi{10.1088/0004-637X/773/1/49}

\bibitem[{{Paxton} {et~al.}(2011){Paxton}, {Bildsten}, {Dotter}, {Herwig},
  {Lesaffre}, \& {Timmes}}]{paxt11}
{Paxton}, B., {Bildsten}, L., {Dotter}, A., {et~al.} 2011, \apjs, 192, 3,
  \dodoi{10.1088/0067-0049/192/1/3}

\bibitem[{{Paxton} {et~al.}(2013){Paxton}, {Cantiello}, {Arras}, {Bildsten},
  {Brown}, {Dotter}, {Mankovich}, {Montgomery}, {Stello}, {Timmes}, \&
  {Townsend}}]{paxt13}
{Paxton}, B., {Cantiello}, M., {Arras}, P., {et~al.} 2013, \apjs, 208, 4,
  \dodoi{10.1088/0067-0049/208/1/4}

\bibitem[{{Paxton} {et~al.}(2015){Paxton}, {Marchant}, {Schwab}, {Bauer},
  {Bildsten}, {Cantiello}, {Dessart}, {Farmer}, {Hu}, {Langer}, {Townsend},
  {Townsley}, \& {Timmes}}]{paxt15a}
{Paxton}, B., {Marchant}, P., {Schwab}, J., {et~al.} 2015, \apjs, 220, 15,
  \dodoi{10.1088/0067-0049/220/1/15}

\bibitem[{{Paxton} {et~al.}(2018){Paxton}, {Schwab}, {Bauer}, {Bildsten},
  {Blinnikov}, {Duffell}, {Farmer}, {Goldberg}, {Marchant}, {Sorokina},
  {Thoul}, {Townsend}, \& {Timmes}}]{paxt18a}
{Paxton}, B., {Schwab}, J., {Bauer}, E.~B., {et~al.} 2018, \apjs, 234, 34,
  \dodoi{10.3847/1538-4365/aaa5a8}

\bibitem[{{Paxton} {et~al.}(2019){Paxton}, {Smolec}, {Schwab}, {Gautschy},
  {Bildsten}, {Cantiello}, {Dotter}, {Farmer}, {Goldberg}, {Jermyn}, {Kanbur},
  {Marchant}, {Thoul}, {Townsend}, {Wolf}, {Zhang}, \& {Timmes}}]{paxt19a}
{Paxton}, B., {Smolec}, R., {Schwab}, J., {et~al.} 2019, \apjs, 243, 10,
  \dodoi{10.3847/1538-4365/ab2241}

\bibitem[{{Ruiter} {et~al.}(2009){Ruiter}, {Belczynski}, \&
  {Fryer}}]{Ruiter2009}
{Ruiter}, A.~J., {Belczynski}, K., \& {Fryer}, C. 2009, \apj, 699, 2026,
  \dodoi{10.1088/0004-637X/699/2/2026}

\bibitem[{{Schwab} {et~al.}(2016){Schwab}, {Quataert}, \& {Kasen}}]{Schwab2016}
{Schwab}, J., {Quataert}, E., \& {Kasen}, D. 2016, \mnras, 463, 3461,
  \dodoi{10.1093/mnras/stw2249}

\bibitem[{{Schwab} \& {Garaud}(2019)}]{schw19a}
{Schwab}, J., \& {Garaud}, P. 2019, \apj, 876, 10

\bibitem[{{Shen} {et~al.}(2012){Shen}, {Bildsten}, {Kasen}, \&
  {Quataert}}]{Shen12}
{Shen}, K.~J., {Bildsten}, L., {Kasen}, D., \& {Quataert}, E. 2012, \apj, 748,
  35, \dodoi{10.1088/0004-637X/748/1/35}

\bibitem[{{Shen} {et~al.}(2018){Shen}, {Kasen}, {Miles}, \&
  {Townsley}}]{Shen2018}
{Shen}, K.~J., {Kasen}, D., {Miles}, B.~J., \& {Townsley}, D.~M. 2018, \apj,
  854, 52, \dodoi{10.3847/1538-4357/aaa8de}

\bibitem[{{Shen} {et~al.}(2017){Shen}, {Toonen}, \& {Graur}}]{Shen17}
{Shen}, K.~J., {Toonen}, S., \& {Graur}, O. 2017, \apjl, 851, L50,
  \dodoi{10.3847/2041-8213/aaa015}

\bibitem[{{Takaro} {et~al.}(2020){Takaro}, {Foley}, {McCully}, {Fong}, {Jha},
  {Narayan}, {Rest}, {Stritzinger}, \& {McKinnon}}]{Takaro2020}
{Takaro}, T., {Foley}, R.~J., {McCully}, C., {et~al.} 2020, \mnras, 493, 986,
  \dodoi{10.1093/mnras/staa294}

\bibitem[{{Willcox} {et~al.}(2016){Willcox}, {Townsley}, {Calder},
  {Denissenkov}, \& {Herwig}}]{wilc16a}
{Willcox}, D.~E., {Townsley}, D.~M., {Calder}, A.~C., {Denissenkov}, P.~A., \&
  {Herwig}, F. 2016, \apj, 832, 13

\bibitem[{{Wong} \& {Schwab}(2019)}]{Wong2019}
{Wong}, T. L.~S., \& {Schwab}, J. 2019, \apj, 878, 100,
  \dodoi{10.3847/1538-4357/ab1b49}

\bibitem[{{Wong} {et~al.}(2021){Wong}, {Schwab}, \& {G{\"o}tberg}}]{Wong2021}
{Wong}, T. L.~S., {Schwab}, J., \& {G{\"o}tberg}, Y. 2021, \apj, 922, 241,
  \dodoi{10.3847/1538-4357/ac27ae}

\bibitem[{{Yoon} \& {Langer}(2003)}]{Yoon03}
{Yoon}, S.~C., \& {Langer}, N. 2003, \aap, 412, L53,
  \dodoi{10.1051/0004-6361:20034607}

\end{thebibliography}
\end{document}